\documentstyle[11pt,a4,epsf]{article}
\newcommand{\ee}{\end{equation}}
\newcommand{\be}{\begin{equation}}
\newcommand{\bef}{\begin{figure}}
\newcommand{\eef}{\end{figure}}
\newcommand{\hmp}{ h^{-1}Mpc}
\newcommand{\etal}{{\it et al.}}
\def\spose#1{\hbox to 0pt{#1\hss}}
\def\ltapprox{\mathrel{\spose{\lower 3pt\hbox{$\mathchar"218$}}
 \raise 2.0pt\hbox{$\mathchar"13C$}}}
\def\gtapprox{\mathrel{\spose{\lower 3pt\hbox{$\mathchar"218$}}
 \raise 2.0pt\hbox{$\mathchar"13E$}}}
\def\inapprox{\mathrel{\spose{\lower 3pt\hbox{$\mathchar"218$}}
 \raise 2.0pt\hbox{$\mathchar"232$}}}

\begin{document}
\centerline{\LARGE Proceedings of the Workshop}
\centerline{}
\centerline{}
\centerline{\Large "Astro-particle physics"}
\centerline{}
\centerline{\Large Ringberg Castle 15.10-19.10 1996}
\centerline{}
\centerline{\Large F. Sylos Labini, L. Pietronero and M. Montuori}
\centerline{}
\centerline{INFM Sezione di Roma1, and} 
\centerline{ Dipartimento di Fisica, Universit\`a di Roma
``La Sapienza''}
\centerline{ P.le A. Moro 2, I-00185 Roma, Italy.}

\bigskip

\section{Introduction}

In this lecture we will try to address the "frequently asked questions 
about fractals"
in the field of large scale galaxy distribution. This paper takes its origin
from a very interesting discussion we
 had at this meeting. A lot of points were raised, 
and we try 
to make clear the fundamentals ones. For a more detailed discussion we refer the 
reader to \cite{cp92} \cite{bslmp94} for a basic introduction to this 
approach, and to \cite{pmsl97} and 
\cite{slmp97} for a review on the more recent results. 

In Sec.2, we briefly introduce the basics concepts 
of fractal geometry and the 
methods of correlation analysis, that are usually used in Statistical Mechanics.
Moreover we present the results of our analysis in the case of 
real galaxy and cluster three dimensional samples. 
In Sec.3 we discuss the 
main points that have been raised during the "Ringberg discussion".
Finally in Sec.4 we summarize our main conclusions.

\section{General properties of fractal distributions}

A fractal consists of a system in which more
and more structures appear at smaller and
smaller scales and the structures at small
scales are similar to the one at large scales.
 Starting from a
 point occupied by an object we count how 
many objects are present within a volume 
characterized by a certain length scale in
 order to establish a generalized "mass-length" 
relation from which one can define the fractal 
dimension.
   We can then write a relation
 between $\:N$ ("mass") and $\:r$ ("length") of type \cite{man83}:
\be
\label{l2}
N(r) = B\cdot r^{D}
\ee
where the fractal dimension is $D$ and the prefactor $\:B$ 
is instead related to the lower cut-offs .
It should be noted that Eq.\ref{l2} corresponds 
to a smooth convolution of a strongly 
 fluctuating function. Therefore a fractal 
 structure is always connected with large
 fluctuations and clustering at all scales.
 From Eq.\ref{l2} we can readily compute the 
 average density $\:<n>$ for a spherical sample of
 radius $\:R_{s}$ which contains a portion 
 of the fractal structure: 
\be
\label{l5}
<n> =\frac{ N(R_{s})}{V(R_{s})} = \frac{3}{4\pi } B R_{s}^{-(3-D)}
\ee
From Eq.\ref{l5} we see that the average density 
is not a meaningful concept in a fractal 
because it depends explicitly on the sample 
size $\:R_{s}$. We can also see that for 
$\:R_{s} \rightarrow \infty$ the average density 
$\:<n> \rightarrow  0 $, 
therefore a fractal structure is asymptotically  
dominated by voids..

It is useful to introduce the conditional density from an point 
occupied as:
\be
\label{l6}
\Gamma (r)= S^{-1}\frac{ dN(r)}{dr} = \frac{D}{4\pi } B r ^{-(3-D)}
\ee
where $\:S(r)$ is the area of  a spherical shell of radius $\:r$.
Usually the exponent that defines the decay
 of the conditional density $\:(3-D)$ is called 
the codimension and it corresponds to the 
exponent $\:\gamma$ of the galaxy distribution.
 
We can now describe  how to perform the correct correlation
analysis that can be applied in the case 
of an irregular distribution as well as of a regular one. 
We may start recalling  the 
concept of correlation. If the presence of an object at the point $r_1$ 
influences the probability of finding another object 
at $r_2$, 
these two points are correlated. Therefore there is a correlation
at  $r$ if, on average
\be
\label{e324}
G(r) = \langle n(0)n(r)\rangle   \ne \langle n\rangle  ^2.
\ee
where we average on all occupied points chosen as origin.
On the other hand there is no correlation if
\be
\label{e325}
G(r) \approx \langle n\rangle  ^2.
\ee
The physically meaningful definition of  $\lambda_0$ 
is therefore the length scale which separates correlated regimes from
uncorrelated ones.

In practice, it  is useful 
to normalize the correlation function (CF) 
of Eq.\ref{e324}
to the size  of the
sample under analysis. Then we use, following \cite{cp92}
\be
\label{e326}
\Gamma(r) = \frac{<n(r)n(0)>}{<n>} = \frac{G(r)}{<n>}
\ee
where $\:<n>$ is the average density of the sample.  We stress
that this normalization does not introduce any bias even if the average
density is sample-depth dependent,
as in the case of fractal distributions,
because it represents
only an overall normalizing factor. 
In order to compare results from different catalogs
it is however more useful to use $\Gamma(r)$, in which
the size of a catalog only appears via the combination
$N^{-1}\sum_{i=1}^{N}$ so that a larger sample 
volume only enlarges the statistical sample over which averages are taken.
 $G(r)$ instead 
has an amplitude that is an explicit function of the sample's size
scale.

The CF of Eq.\ref{e326}
can be computed by the following expression
\be
\label{e327}
\Gamma(r) = \frac{1}{N} \sum_{i=1}^{N} \frac{1}{4 \pi r^2 \Delta r}
\int_{r}^{r+\Delta r} n(\vec{r}_i+\vec{r'})d\vec{r'} = 
\frac{BD}{4 \pi} r^{D-3}
\ee
where the last equality follows from Eq.\ref{l6}.
This function measures the average density at distance $\:\vec{r}$ from an
occupied point at $\vec{r_i}$
and it is called the {\it conditional density} \cite{cp92}.
If the distribution is fractal up to a certain distance $\lambda_0$,
and then it becomes homogeneous, we have that
$$
\Gamma(r) = \frac{BD}{4 \pi} r^{D-3} \;  r < \lambda_0
$$
\be
\label{e327b} 
\Gamma(r)= \frac{BD}{4 \pi} \lambda_0^{D-3} \; r \geq \lambda_0
\ee

It is also very useful to use the conditional average density
\be
\label{e328}
\Gamma^*(r) = \frac{3}{4 \pi r^3} \int_{0}^{r} 4 \pi r'^2 \Gamma(r') dr' =
\frac{3B}{4 \pi} r^{D-3} 
\ee
This function would produce an artificial smoothing of
rapidly varying fluctuations, but it correctly
reproduces global properties   \cite{cp92}.

For a fractal structure, $\Gamma(r)$ has a power law behaviour
and the conditional average density $\Gamma^*(r)$ has the form
\be
\label{e329}
\Gamma^*(r)= \frac{3}{D} \Gamma(r)
\ee
for an homogenous distribution ($D=3$) these two functions
are exactly the same and equal to the average density.

\subsection{The $\xi(r)$ correlation function for a fractal }

  Pietronero  and collaborators \cite{pie87} \cite{cps88} 
\cite{cp92} 
have clarified some crucial points of the
standard correlations analysis, and in particular they have discussed the 
physical meaning
of the so-called {\it "correlation length"}
  $\:r_{0}$ found with the standard
approach \cite{pee80} \cite{dp83} and defined by the relation:
\be
\label{e330}
\xi(r_{0})= 1
\ee
where
\be
\label{e331}
\xi(r) = \frac{<n(\vec{r_{0}})n(\vec{r_{0}}+ \vec{r})>}{<n>^{2}}-1
\ee
is the two point correlation function used in the standard analysis.
The basic point that \cite{cp92} stressed,
is that the mean density, $<n>$,
used in the normalization of $\:\xi(r)$, is not a well defined quantity
in the case
of self-similar distribution and it is a direct function of the sample size.
Hence only in the case that
the homogeneity  has been reached well within the sample
limits the $\:\xi(r)$-analysis is meaningful, otherwise
the a priori assumption of homogeneity is incorrect and the
characteristic lengths, like $\:r_{0}$, became spurious.

 For example, following \cite{cp92}
 the expression of the $\:\xi(r)$ in the case of
 fractal distributions is:
\be
\label{e332}
\xi(r) = ((3-\gamma)/3)(r/R_{s})^{-\gamma} -1
\ee
where $\:R_{s}$ is the depth of the spherical volume where one computes the
average density from Eq.\ref{l5}.
From Eq.\ref{e332} it follows that

i.) the so-called correlation
length $\:r_{0}$ (defined as $\:\xi(r_{0}) = 1$)
is a linear function of the sample size $\:R_{s}$
\be
\label{e333}
r_{0} = ((3-\gamma)/6)^{\frac{1}{\gamma}}R_{s}
\ee
and hence it is a spurious quantity without  physical meaning but it is
simply related to the sample finite size.

ii.) $\:\xi(r)$ is power law only for
\be
\label{e334}
((3-\gamma)/3)(r/R_{s})^{-\gamma}  >> 1
\ee
hence for $r \ll r_0$: for larger distances there is a clear deviation
from the power law behaviour due to the definition of $\xi(r)$.
This deviation, however, is just due to the size of
 the observational sample and does not correspond to any real change
of the correlation properties. It is clear that if one estimates the
 exponent of $\xi(r)$ at distances $r \ltapprox r_0$, one
 systematically obtains a higher value of the correlation exponent
 due to the break of $\xi(r)$ in the log-log plot.

The analysis
 performed by $\xi(r)$ is therefore mathematically inconsistent, if
 a clear cut-off towards homogeneity has not been reached, because
 it gives an information that is not related to the real physical
 features of the distribution in the sample, but to the size of the
sample itself.

\subsection{Analysis of the Galaxy distributions}

One of the most important issues
raised by all the recently  catalogues is that the scale of {\em the
largest inhomogeneities}
 is comparable with {\it the
extent of the surveys}  
in which they are detected. These galaxy catalogues
probe scales from $\: \sim 100-200 h^{-1} Mpc$ 
for the wide angle surveys, up to $\:\sim  1000 h^{-1} Mpc$ 
for the deeper pencil  beam surveys
(that cover a very narrow solid angle)
and show that
the Large Scale Structures (LSS) 
are the characteristic features of the visible matter distribution.
From these data  a new picture  emerges
 in which the 
scale of homogeneity seems to shift to a very large value, not 
still identified.

In the past years 
\cite{slgmp96} \cite{slmp96} \cite{slp96} \cite{sla96}
\cite{dmppsl96} \cite{pmsl97} (see \cite{slmp97}  for a review )
we have analyzed the statistical properties of 
several redshift surveys (see Tab.\ref{tab1})
with the methods of modern Statistical Physics.
\begin{table}
\caption{\label{tab1}The volume 
limited catalogues are characterized by the following 
parameters:
- $R_d (\hmp)$ is the depth of the catalogue
- $\Omega$ is the solid angle
- $R_s (\hmp)$ is the radius of the largest sphere 
that can be contained in the catalogue volume. 
This gives the limit of statistical validity of the sample.
- $r_0(\hmp)$  is the length at which $\xi(r) \equiv 1$.
- $\lambda_0$ is the eventual real crossover to a homogeneous 
distribution that is actually never observed. 
The value of $r_0$ 
is the one obtained 
in the deepest sample. 
The CfA2 and SSRS2 data are not yet available.
 (distance are expressed in $\hmp$).
}
\begin{tabular}{|c|c|c|c|c|c|c|}
\hline
     &      &          &    &              &  &       \\
\rm{Sample} & $\Omega$ ($sr$) & $R_d  $ & $R_s  $ 
& $r_0  $& $D$ & $\lambda_0 $ \\
     &       &    &    &    &  		   &       \\
\hline
CfA1 & 1.83  & 80 & 20 & 6  & $1.7 \pm 0.2$& $>80$    \\
CfA2 & 1.23  & 130& 30 & 10 & 2.0          & $ ? $    \\
PP   & 0.9   & 130& 30 & 10 & $2.0 \pm 0.1$& $> 130$  \\
SSRS1& 1.75  & 120& 35 & 12 & $2.0 \pm 0.1$ & $ >120$      \\
SSRS2& 1.13  & 150& 50 & 15 & 2.0          & $?$      \\
Stromlo-APM& 1.3  & 100& 30 & 10 & $2.2 \pm 0.1$& $ > 150$      \\
LEDA & $4 \pi$ & 300& 150& 45 & $2.1 \pm 0.2$& $>150$   \\
LCRS & 0.12  & 500& 18 & 6  & $1.8 \pm 0.2$& $> 500$    \\
IRAS $2 Jy$ & $4 \pi$ & 60 & 30 & 4.5& $2.0 \pm 0.1$& $ > 50$\\
IRAS $1.2 Jy$ & $4 \pi$ & 80 & 40 & 6& $2.0 \pm 0.1$& $> 50$\\
ESP  & 0.006 & 700& 10 & 5  & $1.9 \pm 0.2$& $>800$   \\
     &       &         &       &      &  &       \\
\hline
\end{tabular}
\end{table}
The main data of our correlation analysis 
are collected in Fig.\ref{fig1} (left part)
 in which we report the 
{\it conditional density as a function of scale}
 for the various catalogues. 
The relative position of the various lines is not arbitrary but it is fixed 
by the luminosity function, a part for the cases of 
IRAS and SSRS1 for which this is 
not possible. The properties derived from different 
catalogues are compatible with each other and show a {\it power law 
decay} for the conditional density from $1 \hmp$ to $150 \hmp$
 without any tendency towards homogenization (flattening). This 
implies necessarily that the value of $r_0$ (derived from the $\xi(r)$ 
approach) will scale with the sample size $R_s$ as shown also from the 
specific data about $r_0$ of the various catalogues. The 
behaviour 
observed  corresponds to a fractal structure with dimension $D \simeq 
2$. The smaller value of CfA1 was due to its limited size. An 
homogeneous distribution would correspond to a flattening of the 
conditional density which is never observed
\bef
 \epsfxsize 12cm 
\centerline{\epsfbox{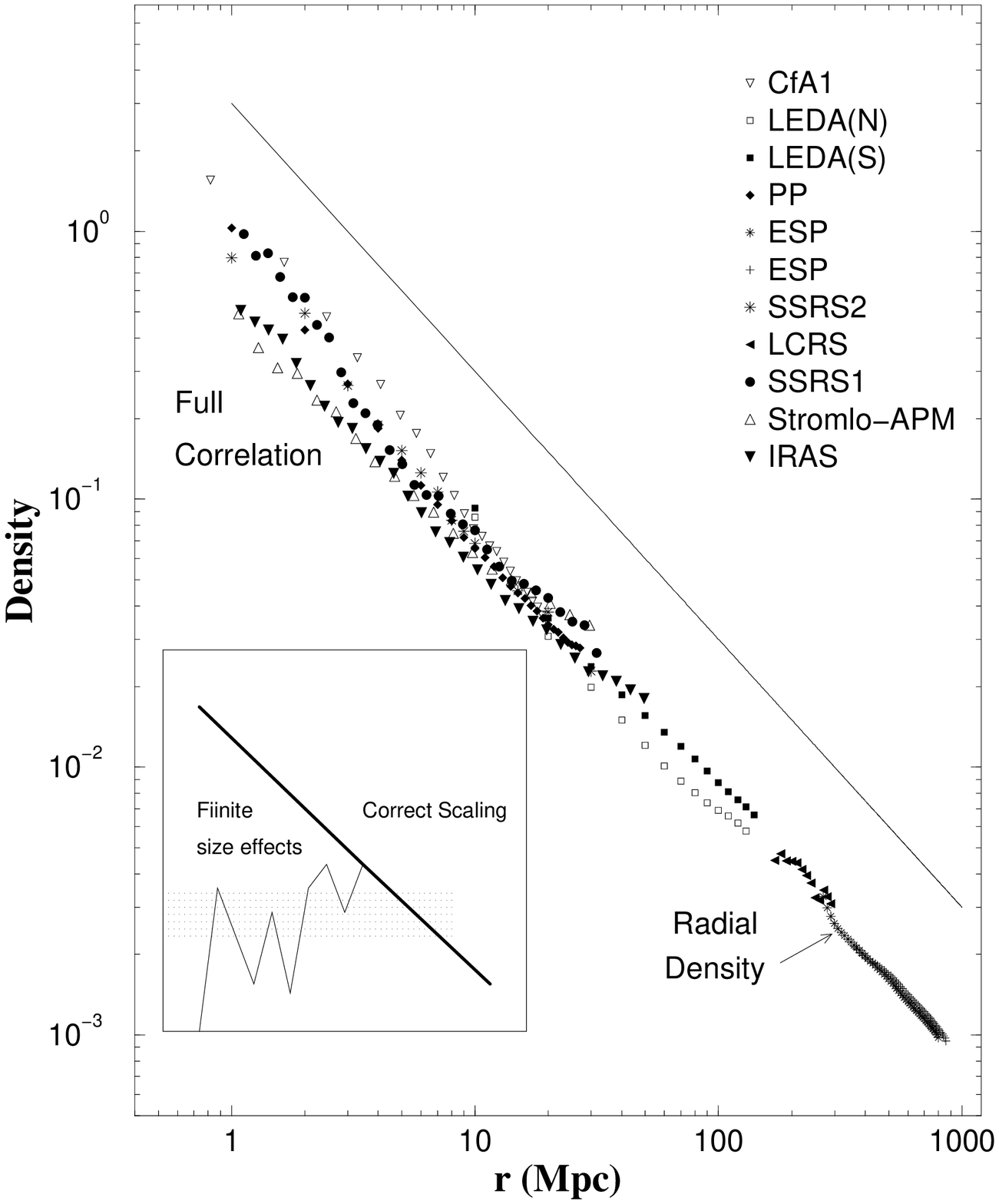}}  
\caption{\label{fig1} Full correlation analysis for the various
 available redshift surveys in the range of distance $0.5 \div 1000
 \hmp$. A reference line with slope $-1$ is also shown, that 
corresponds to fractal dimension $D = 2$. 
 } 
\eef 
It is
 remarkable to stress that the amplitudes and the slopes of the
 different surveys match quite well. From this figure we conclude
 that galaxy correlations show very well defined fractal properties
 in the entire range $0.5 \div 1000 \hmp$ with dimension $D = 2 \pm
 0.2$. Moreover all the surveys are in agreement with each other.

It is interesting to compare the analysis of Fig.\ref{fig1} with 
the usual one, made with the function $\xi(r)$, for the same 
galaxy catalogs. This is reported in Fig.\ref{fig2}
\bef
\epsfxsize 11cm 
\centerline{\epsfbox{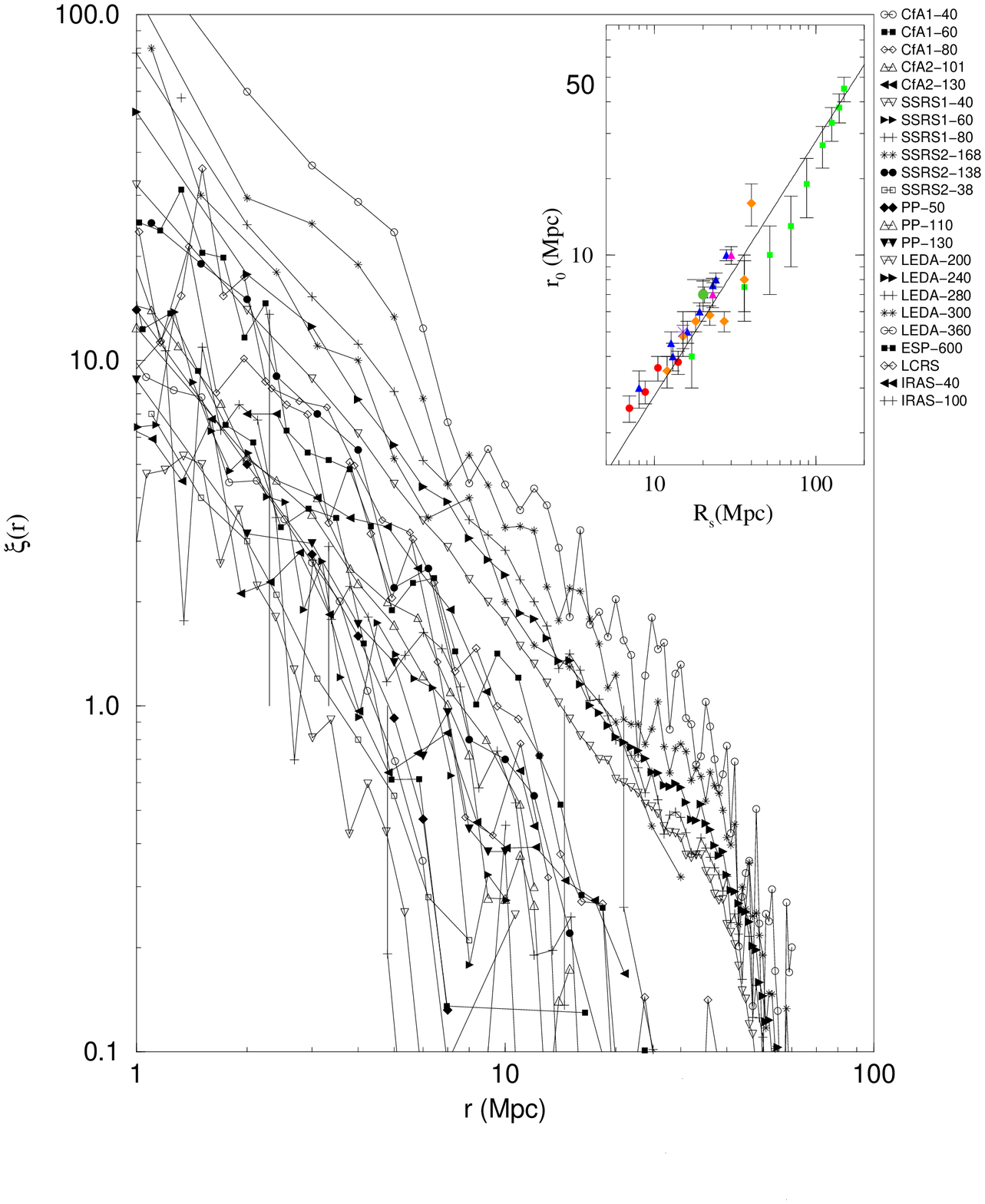}}  
\caption{\label{fig2}
Traditional analyses based on the function $\xi(r)$
of the same galaxy catalogs of the previous figure.
 The usual 
analysis is based on the a priori untested assumptions of 
analyticity and homogeneity. These properties
are not present in the real galaxy distribution and 
the results appear therefore rather confusing. 
This lead to the impression that galaxy catalogs are not good
enough and to a variety of theoretical problems like the 
galaxy-cluster mismatch, luminosity segregation, linear and 
non-linear evolution, etc.. This situation changes completely and 
becomes quite clear if one adopts the more 
general conceptual framework that is at the basis 
the previous figure}
\eef 
and, from this point of view, the various data 
the various data appear to be in strong disagreement with 
each other. This is due to the fact that the usual analysis
looks at the data from the prospective of analyticity and large
scale homogeneity (within each sample). These properties have never
been tested and they are not present in the real galaxy
distribution so the result is rather confusing (Fig.\ref{fig2}).
Once the same data are analyzed with a broader perspective the
situation becomes clear (Fig.\ref{fig1}) and the data of 
different catalogs result in agreement with each other. It is 
important to remark that analyses like those of Fig.\ref{fig2}
have had a profound influence in the field in various ways: 
first the different catalogues appear in conflict with each other.
This has generated the concept of {\it not fair samples} and a 
strong mutual criticism about the validity of the data 
between different authors. In the other cases the
discrepancy observed in Fig.\ref{fig2} have been 
considered real physical problems for which various technical
approaches have been proposed. These problems are, for example, 
the galaxy-cluster mismatch, luminosity segregation, 
the richness-clustering relation and 
 the linear non-linear evolution of the perturbations 
corresponding to the {\it "small"} or  {\it "large"}
amplitudes of fluctuations. We can now see that all this problematic
situation is not real and it arises only from a 
statistical analysis based on inappropriate and too restrictive
 assumptions that do not find any correspondence in the 
physical reality. It is also important to note that,
even if the galaxy distribution would eventually became
homogeneous at larger scales, the use of the above statistical
concepts  is anyhow inappropriate for the range of scales 
in which the system shows fractal correlations as those 
shown in Fig.\ref{fig1}.

  \section{Main points raised in the debate with Dr. Martinez}

\begin{itemize}


\item{\bf Treatment of boundary conditions and weighting schemes}

Given a certain sample of solid angle $\Omega$ and depth $R_d$,
it is important to define which is 
 the maximum distance up to which it 
is possible to compute the correlation function ($\Gamma(r)$ or $\xi(r)$). 
As discussed in \cite{cp92}
(see also \cite{slmp96} \cite{dmppsl96} \cite{slgmp96}),
 we have limited our analysis to an
effective  depth
$R_{s}$ that is of the order of the radius of the maximum
sphere fully contained in the sample volume.
For a catalog with the limits, for example, in right ascension
($\alpha_1 \leq \alpha \leq \alpha_2$) and declination
($\delta_1 \leq \delta \leq \delta_2$) we have that 
\be
\label{ws1}
R_{s} = \frac{R_d sin (\delta \theta /2)}{1+sin(\delta \theta/2)}
\ee
where $\delta \theta = min (\alpha_2 - \alpha_1, \delta_2-\delta_1)$.
In such a way that we eliminate from the statistics
the points for which a sphere of radius {\it r} is not
fully included within the sample boundaries.
Hence we do not make use of any weighting scheme
with the advantage that
we do not make any assumption in the treatment of
the boundaries conditions.
Of course in
doing this, we have a smaller number of points
and we stop our analysis  at a  smaller depth than that
of other authors.
In Tab.\ref{tab1} we report the values of $\Omega$, $R_s$ and $R_d$
for the various catalogs. We can see that, althought LCRS or ESP 
are very deep, the value of $R_s$ is of order of the one of CfA1, and this
is the reason why the value of $r_0$ is almost the same in these 
different surveys. On the other hand, in CfA2 the value of $r_0$ has been 
measured to be $r_0 \approx 11 \hmp$ \cite{par94} and 
in SSRS2 $r_0 \approx 15 \hmp$ \cite{ben96}, because 
their solid angle is quite large. 

 The reason why
$\Gamma(r)$ (or $\xi(r)$) cannot
be computed for $r > R_{s}$
is essentially the following.
 When one evaluates the correlation
function
(or power spectrum \cite{sla96}) beyond $R_{s}$,
then one  makes explicit assumptions on what
lies beyond the sample's boundary.  In fact, even in absence of
corrections for selection effects, one
is forced to consider incomplete shells
calculating $\Gamma(r)$ for $r>R_{s}$,
thereby
implicitly assuming that what one  does not see in the part of the
shell not included in the sample is equal to what is inside (or other
similar weighting schemes).
In other words, the standard calculation
introduces a spurious homogenization which we are trying to remove.

If one could reproduce via an analysis that uses weighting schemes, the 
correct properties of the distribution under analysis, it would be
not necessary to produce wide angle survey, and from a single pencil beam
deep survey it would be possible to study the entire matter
distribution up to very deep scales. It is evident that
this could not be the case.

By the way, we have done  a test \cite{slmp97} 
on the homogenization effects
of weighting schemes on artificial distributions as well as on
real catalogs, finding that the flattening of the 
conditional density
 is indeed introduced owing to  the weighting,
and does not correspond to any real feature in the galaxy distribution.

\item{\bf Luminosity segregation}

A possible explanation of the shift of $r_0$ is based
on the luminosity segregation effect
\cite{dav88} \cite{par94} \cite{ben96}.
We briefly illustrate this approach.
The fact that the giant galaxies are more clustered than
the dwarf ones,
i.e. that they are located in the peaks of the density field,
has given rise to the proposition
that larger objects may correlate up to larger
length scales and that the amplitude of the $\:\xi(r)$ is larger
for giants than for dwarfs one. The deeper VL subsamples
contain galaxies that are in average
brighter than those in the VL subsamples with
smaller depths. As the brighter galaxies should have
a larger correlation length the behavior found in Fig.\ref{fig1}
could be related, at least partially,
with  the phenomenon of luminosity segregation.

We would like to stress that as long as $\Gamma(r)$ has a
 power law decay, $r_0$ {\it 
must be a linear fraction of the sample size} (see Eq.\ref{e333}), and that 
as far as a clear crossover towards homogeneity has been not identified, 
the "correlation length" $r_0$ has no physical
 meaning, just being related to the size of the sample.
{\it Moreover the authors (e.g. \cite{dav88} \cite{par94} \cite{ben96}) 
that have introduced the 
concept do not present {\it any quantitative argument} 
that explains the shift of $r_0$ with sample size.} In addition
as we have discussed previsoulsy, as far as a clear cut-off 
towards homogeneity has not been identified, the analysis performed
by the $\xi(r)$ function is misleading, i.e. it does not give a 
physically meaningful result. 

We have discussed in detail in \cite{slp96} 
that the observation
that the giant galaxies are more clustered
than the dwarf ones, i.e. that the massive elliptical
galaxies lie
in the peaks of the density field, is a consequence of the
self-similar behavior of the whole matter distribution. The increasing
of the correlation length of the $\xi(r)$ has
nothing to do with this effect \cite{cp92} \cite{bslmp94}.

Finally we would like to stress the conceptual problems of the 
interpretation of the scaling of $r_0$ by the luminosity segregation phenomenon.
Suppose we have two kind of galaxies of different masses, one of type $A$
and the other of type $B$. Suppose for simplicity that the mass of 
the galaxies of type $A$ is twice that of the $B$. 
The proposition 
"galaxies of different luminosities (masses) correlate in different 
ways" implies that the gravitational interaction is able to distinguish
between a situation in which there is, in a certain place, a galaxy of type $A$
or two galaxies of type $B$ placed very near. This seem to be not possible
as the gravitational interaction is due the sum of all the masses.

We can go farther by showing
 the inconsistency of the proposition "galaxies
of different luminosities have a different correlation length". Suppose that
the galaxies of type $A$ have a smaller correlation length
than that of the galaxies of type $B$ 
This means that the galaxies of type $B$ are still correlated (in terms of 
the conditional density) when the galaxies of type $A$ are 
homogeneously distributed. This means that the galaxies of type $A$ 
should fill the voids of galaxies of type $B$. This is not the case, as 
the voids
are empty of all types of galaxies, and it seems that the large scale structures
distribution is independent
 on  the galaxy morphological types.

\item{\bf Change of slope at small scale: is it a real physical effect ?}

The conditional  density $\Gamma(r)$ (Eq.\ref{l6} and Eq.\ref{e327}) measures 
the density in a shell of thickness $\Delta r$ at distance $r$ from an 
occupied point, and then it is averaged over all the points of the sample. 
In practice, we have three possibilities
for $\Gamma(r)$: i) $D=3$: in this case this function is
simply a constant. ii) $0<D<3$ In this case the conditional density has a power law
decay  with exponent $-\gamma=D-3$. Finally iii) $D=0$: this is the limiting case 
in which there are no further points in the sample except the observer. In such 
a situation we have that $\Gamma(r)$ has a $1/r^3$, and scales as the three 
dimensional volume.

Suppose now, for simplicity,  we have a spherical sample of volume $V$ in which there 
are $N$ points, and we want to measure the conditional density.
The maximum depth is limited by the radius of the sample (as previously discussed), 
while the minimum depth 
depends on the number of points contained in the volume. 
For a Poisson distribution the mean average distance between near neighbor 
is of the order $\ell \sim (V/N)^{\frac{1}{3}}$. Strictly speaking such a 
relation does not holds in the case of a fractal distribution, 
but it gives just an order of magnitude of
the distance between near neighbor.
 If we measure the conditional density at distances $ r \ll \ell$, 
we are biased by a {\it finite size effect}: due the depletion of points at these distances 
we will underestimate the real conditional  density finding an higher value 
for the correlation exponent (and hence a lower value for the fractal dimension). 
In the limiting case that we find no points in the range $ l_{min} \ltapprox r 
\ltapprox \ell$ the slope will be $\gamma=-3$, that corresponds to 
$D=0$. In general, when one  measures $\Gamma(r)$ at distances that correspond to 
a fraction of $\ell$, one finds systematically an higher value of the 
conditional density exponent. 
This is completely spurious and due to the depletion of
points at such distances. 

For example in a real survey, in order to check this effect, one should measure 
$\Gamma(r)$ in volume 
limited samples with different values of $\ell$. In general we find
that the more sparse  samples exhibit a change of slope a small distances, while
for the samples 
 for which $\ell$ is quite small, the change of slope at small distances 
is found (e.g. 
\cite{slmp96}). 
In general for a typical sample of galaxies $\ell \sim 2 \div 10 \hmp$, so that 
the behaviour of $\Gamma(r)$ at distances of some Megaparsec is generally affected by this 
finite size effect. A way to reduce this effect is to choose
 properly the thickness 
$\Delta r$ of the shell in which the conditional density is computed: this means that 
at small distances $\Delta r$ must be of the order of $\ell$ and not 
smaller than this value. In general we have found that best way to optimize this 
estimation is to choose logarithm interval for $\Delta r$, as a function of the scale.

\item{\bf Statistical stability of the correlation analysis}

To check that possible  errors in the apparent magnitude
do not affect seriously the behavior of $\Gamma(r)$
one can  perform the various tests. For example
one can  change the apparent magnitude
of galaxies in the whole catalog by a random factor $\delta m$
with $\delta m= \pm 0.2, 0.4, 0.6, 0.8$ and $1$. We find that the
number of galaxies in the VL samples change from $5 \%$ up to
$15 \%$ and that the amplitude and the slope of $\Gamma(r)$ are
substantially stable and there are not any significant changes
in their behavior. This is because $\Gamma(r)$ measures a global quantity that
is very robust with respect to these possible errors.
There are several other tests that one can perform, and that 
are discussed in detail in \cite{slmp97}. 
However would like to stress, that a
fractal distribution has a very strong property: it shows
 power law
correlations up to the sample depth. Such correlations {\it cannot be due}
neither 
by an inhomogeneous sampling of an homogeneous distribution, 
nor by some selection effects that may occur in the observations. 
Namely, suppose that  a certain kind of sampling reduces the 
number of galaxies as a function of distance \cite{dav97}.
 Such an effect in no way can lead to long range correlations,
because when one computes $\Gamma(r)$, one makes an average 
over all the points inside the survey.

\item{\bf Samples validity and dilution effects: what is a fair sample ?}

How many galaxies one needs in order   to characterize correctly
(statistically) 
 the large scale distribution of visible matter ? This fundamental
 question is addressed in this section, and it will allows us to
 understand some basic properties of the statistical analysis of 
 galaxy surveys. In such a way, we will be able to clarify  the
 concept of "fair sample", i.e. a sample that contains a
 statistically meaningful information \cite{slgmp96} \cite{slmp97}

  We have discussed in the pervious sections the properties of
 fractal structures and in particular we have stressed  the
 intrinsic highly fluctuating nature of such distributions. In this
 perspective it is important to clarify the concept of  {\it "fair
 sample"}. Often this concept is used as synonymous of a
 homogeneous sample (see for example \cite{dac94}). 
  So the analysis of catalogues along the traditional
 lines often leads to the conclusion that we still do not have a
 fair sample and deeper surveys are needed to derive the correct
 correlation properties. A corollary of this point of view is that
 since we do not have a fair sample its statistical analysis cannot
 be taken too seriously.

 This point of view is highly misleading because we have seen
(Sec.1)  that self-similar structures never become homogeneous, so
 any sample containing a self-similar (fractal) structure would
 automatically be declared "not fair" and therefore impossible to
 analyze. The situation is actually much more interesting otherwise
 the statistical mechanics of complex systems would not exist.
 Homogeneity is a property and not a condition of statistical
 validity of the sample. A non homogeneous system can have well
 defined statistical properties in terms of scale invariant
 correlations, that may be perfectly well defined. The whole
 studies of fractal structures are about this \cite{pt86} \cite{epv95}.
 Therefore one should distinguish
 between a "statistical fair sample", which is a sample in which
 there are enough points to derive some statistical properties
 unambiguously and a homogeneous sample, that is a property that can
 be present or not but that has nothing to do with the statistical
 validity of the sample itself. We have seen in Sec.3  
 that even the small
 sample like CfA1 is statistically fair up to a distance that can
 be defined unambiguously (i.e $\sim 20 \hmp$).

In \cite{slgmp96} we have  studied the following 
 question. Given a sample with a well defined volume, which is the
 minimum number of points that it should contain in order to have
 a {\em statistically  fair} sample, even if one computes averages
 over all the points, such as the conditional density and the
 conditional  average density ?  

Suppose that the
sample volume is a portion of a sphere with a solid angle $\Omega$ and radius
$R$, the mass ($N(<R)$) length ($R$) relation can be written as
\cite{cp92}  
\begin{equation}
\label{prl1}
N(<R) = B\left(\frac{\Omega}{4\pi} \right) R^D  
\end{equation}
where $D$ is the fractal dimension or, for the homogeneous case, $D=3$. 
The prefactor   $B$ is a constant and it is 
related to the lower cut-off of the fractal 
structure \cite{cp92} \cite{slgmp96}.
In this letter we consider  galaxies of different luminosities 
having the 
same clustering properties (i.e. equal fractal dimension):
this is a crude approximation and the more complex situation can be
described in terms of multifractal \cite{slp96}.

In principle  Eq.\ref{prl1} should refer  to {\it all}
the galaxies existing in a given
volume. If instead we have a VL sample,
 we will see only a fraction
$N_{VL}(R) = p \cdot N(<R)$  (where $p<1$)
 of the total number $N(<R)$. In order to estimate the
fraction $p$ it is necessary to know the luminosity function $\phi(L)$ that
gives the fraction of galaxies whose absolute luminosity ($L$) is between 
$L$ and
$L+dL$ \cite{sc76} 
This function has been extensively measured \cite{dac94} 
and it consists of a
power law extending from a minimal value $L_{min}$ to a maximum value $L^*$
defined by an exponential cut-off. Therefore we can express the fraction $p$ as
\begin{equation}
\label{prl2}
 0< p = \frac{\int_{L_{VL}}^\infty  \phi(L)dL} 
{\int^\infty_{L_{min}} \phi(L)dL} < 1
\end{equation}
where $L_{VL}$ is the minimal absolute luminosity that 
characterizes the VL
sample. The quantity $L_{min}$ is 
the fainter absolute flux (magnitude $M_{lim}$)
surveyed in the catalog (usually $M_{min} \sim -11 \div -12$).

We have performed several tests   in real galaxy samples as well 
as in artificial distributions with a priori assigned properties. In particular 
we have eliminated randomly points from the original 
distribution: such a procedure, 
for the law of codimension additivity \cite{man83} does not change the fractal 
dimension, but only the prefactor in Eq.\ref{prl1}. In such a way we can control,
quantitatively, the behaviour of the conditional density as the value of $p$
decreases, and we 
are able to conclude that, if the system has self-similar properties,
a reliable
 correlation
a\-na\-lysis is only possible if $p$ is substantially larger than 
$1 \div 2\%$ \cite{slgmp96}. 
Below
this value the statistical significance becomes questionable just because
 the sample is
too sparse and large scale correlations are destroyed by this effect.

In relation to the statistical validity it is interesting to consider the 
IRAS catalogues  because they seem to differ from all the other ones and 
to show some tendency towards homogenization at a relatively small scale 
Actually the point of apparent homogeneity 
is only present in some samples, it varies 
from sample to sample between $\sim 15 \div 25 \hmp$ 
and it is strongly dependent on the dilution of the sample.
Considering that structures and voids are much 
larger than this scale and that the IRAS galaxies appear to be just 
where luminous galaxies are it is clear that this tendency appears 
suspicious. One of the characteristic of the IRAS catalogues with respect 
to all the other ones is an extreme degree of dilution: this catalogue 
contains only a very small fraction of all the galaxies. It is important  
therefore to study what happens to the properties of a given sample if 
one dilutes randomly the galaxy distribution up to the IRAS limits. A 
good test can be done by considering the Perseus Pisces catalogue and 
eliminating galaxies from it. The 
original distribution shows a well defined fractal behavior. 
By diluting it to the level of IRAS one observes an artificial flattening 
of the correlations \cite{slgmp96} (see Fig.\ref{fig3}).
\bef
\epsfxsize 14cm
\centerline{\epsfbox{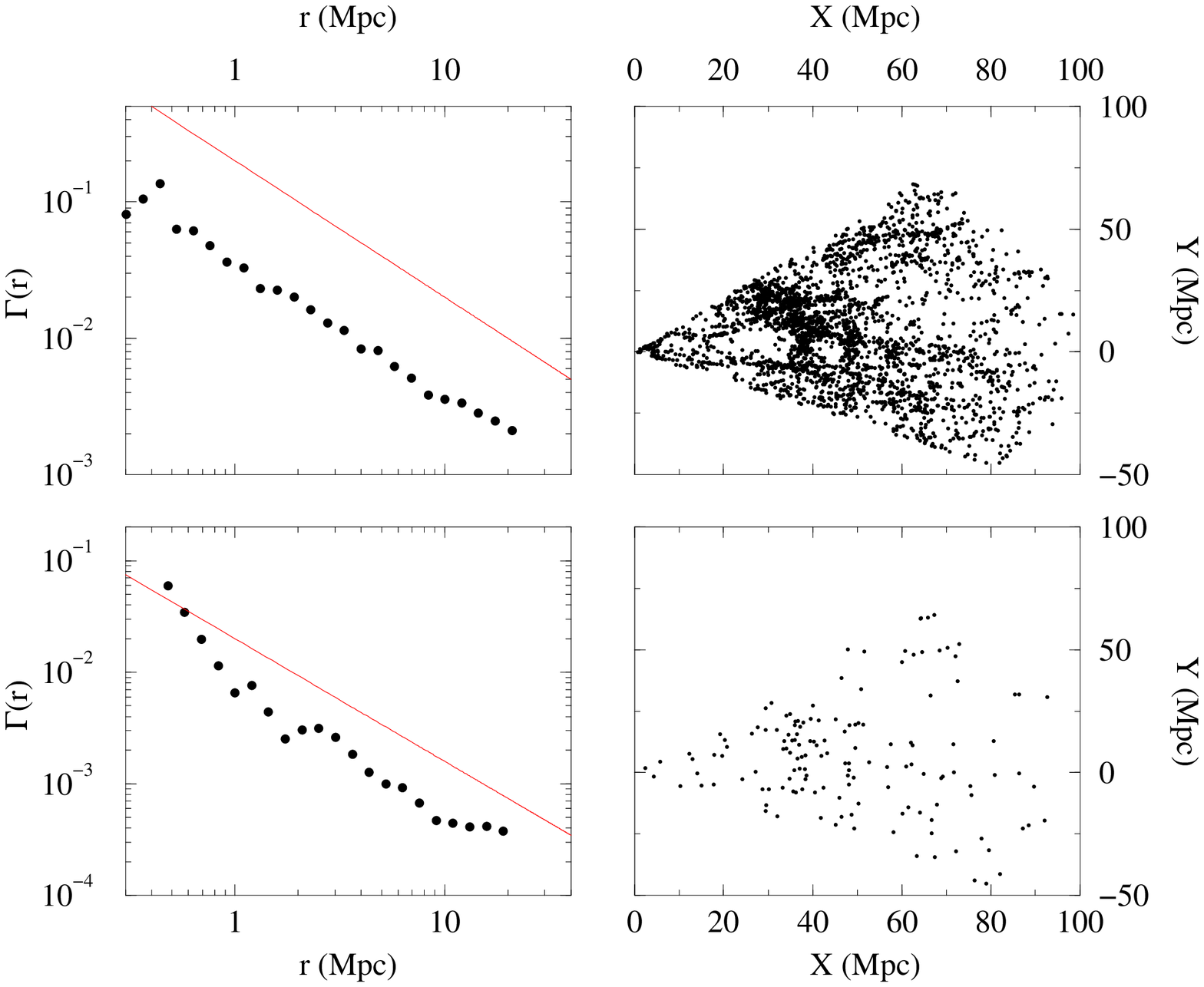}}
\caption{ \label{fig3}
{\it Top panels:} The conditional density (left) for a volume limited sample
of the full 
Perseus-Pisces redshift survey (right). 
The percentage of galaxies present in the sample is $\sim 6 \%$.
The slope is $-\gamma=-1$.
{\it Bottom panels:} In this case the percentage of galaxies is $\sim 1 \%$,
and the number of galaxies is the same of the IRAS $1.2 Jy$ sample 
in the same region of the sky. We can see that at small scale 
we have a $1/r^3$ decay just due to the sparseness of the sample,
while at large scale the shot noise of the sparse sampling 
overcomes the real correlations
and produces an apparent trend to homogenization.
In our opinion, this effect, due to sparseness of 
this sample, is the origin of the apparent trend 
towards homogenization observed in some of the 
IRAS samples.}
\eef
 This effect does not correspond to a real 
homogenization but it is due to the of dilution. In fact it can be 
shown that when the dilution is such that the average distance 
between galaxies becomes comparable with the largest voids 
(lacunarity) of the original structure there is a loss of correlation 
and the shot noise of the sparse sampling 
overcomes the real correlations
and produces an apparent trend to homogenization.
This allows us to reconcile this peculiarity of the 
IRAS data with the properties of all the other catalogues.
Analogous considerations for other sparse samples 
like QDOT and the Stromlo-APM samples \cite{slmp97}.

\item{\bf Problem of the power spectrum analysis}

{\it In the following we argue that both features, bending and scaling,
are a manifestation of 
the finiteness of the survey volume, and that they
cannot be
interpreted as the convergence to homogeneity, nor to a power spectrum
flattening \cite{sla96}.} 
The systematic effect of the survey finite size is in fact
to suppress
power at large scale, mimicking a real flattening.
Clearly, this effect occurs whenever galaxies have  not 
a definite  correlation scale, with respect to the survey.
 We push this argument further, by showing that
even a fractal distribution of matter, 
which never reaches  homogeneity, shows a sharp flattening. Such a 
flattening is  partially corrected, but not quite eliminated,
  when the correction proposed by \cite{pn91} 
 is
applied to the data.
 We show also
 how the amplitude of the
power spectrum depends on the survey size as long as 
the system shows long-range correlations.

The standard  power spectrum (SPS) 
measures directly the contributions of different scales to the galaxy
density contrast $\:\delta\rho/\rho$.
It is clear that the density contrast, 
and all the quantities based on it,  is meaningful only when one can define
a constant density, i.e. reliably identify
the sample density with   
the average density of all the Universe.
In other words in  {\it the SPS analysis 
one assumes that the survey volume is large enough
to contain a  homogeneous sample.} 
When this is not true, and we argue that is 
indeed an incorrect assumption
in all the cases investigated so far, a false interpretation of the results may
occur, since both 
the shape and the amplitude of the power spectrum (or correlation
function) depend on the
survey size.

As we have already mentioned, in a fractal
quantities like 
$\xi(r)$ 
are  scale dependent: in particular both
the amplitude and the shape of $\xi(r)$ are 
therefore scale-dependent  in the
case of a fractal distribution. 
It is clear  that the same kind of 
finite size effects  are also present when computing the SPS, so that 
it is very dangerous to identify real physical features induced
from the SPS analysis without first a firm determination of the
homogeneity scale.

The SPS for a fractal distribution 
inside a sphere of radius $R_s$ is
\be 
\label{e19}
P(k) =  \int^{R_{s}}_{0} 
   4\pi \frac{\sin(kr)}{kr} \left[ \frac{3-\gamma}{3} 
\left(\frac{r}{R_{s}}\right)^{-\gamma} -1\right] r^{2}dr=
\frac{a(k,R_s) R_{s}^{3-D}}{k^{D}}-\frac{b(k,R_s)}{k^{3}}\,.
 \ee
Notice that the integral has to be evaluated inside $R_s$
because we want to compare $P(k)$  with its {\it estimation}
 in a finite size spherical survey of scale $R_s$. 
 In the  general case, we must deconvolve the
 window contribution 
 from $P(k)$; $R_s$ is then a characteristic window scale. 
Eq. (\ref{e19})  shows the two scale-dependent features of the PS. First,
the amplitude of the PS 
depends on the sample depth.
Secondly,
the shape of the PS 
is characterized by two  scaling regimes:
the first one, at high wavenumbers, 
is related to the fractal dimension of the 
distribution in real space, 
while the second one arises only because of 
the finiteness of the sample.

\item{\bf About multifractality}

We now briefly introduce
 the multifractal picture that is
 a refinement and generalization of the fractal properties 
\cite{pv87} \cite{ben84} \cite{cp92} \cite{bslmp94} \cite{slmp96}
\cite{slp96}
arising naturally
in the case of self-similar distributions. If one does not
 consider the mass one has a simple set  given by the galaxy
 positions (that we call the {\it support}  of the measure
 distribution). Multifractality instead becomes interesting and a
 physically  relevant property when one includes the galaxy masses
 and consider the entire matter distribution \cite{pie87} \cite{cp92}.
In this case the measure distribution  is defined by
 assigning  to each galaxy a weight which is  proportional to its
 mass. The question  of the self-similarity versus homogeneity of
 this set can be  exhaustively discussed in terms of the single
 correlation exponent that corresponds to the fractal dimension  of
 the support of the measure distribution.  
Several authors 
 \cite{mj90} instead considered the eventual
 multifractality  of the support itself. However the physical
 implication of  such an analysis is not clear,  and it does not
 add much to the  question above.   

 In the more complex case of MF
 distributions the scaling properties can be different for
 different regions of the system and one has to introduce a
 continuous  set of exponents to characterize the  system  (the
 multifractal spectrum $f(\alpha)$). The discussion     presented in the
 previous sections  was meant to distinguish between homogeneity and
 scale invariant properties;  it is appropriate also in the case of
 a multifractal. In the latter case the correlation functions we
 have considered would correspond to a single exponent  of a
 multifractal spectrum of exponents, but the issue of homogeneity
 versus scale invariance (fractal or multifractal) remains exactly
 the same. 

We have shown \cite{slp96}
that it is possible to 
  frame  the main properties of the galaxy space and 
luminosity  distribution  in a unified scheme, by  using
 the concept of multifractality (MF). In fact, the
 continuous set of exponents $[\alpha,f(\alpha)]$ 
that describes a MF distribution can characterize 
completely  the galaxy distribution when one considers
  the mass (or luminosity) of galaxies in the analysis. In 
this way many observational evidences are linked together
 and arise naturally from the self-similar properties of the 
distribution.   Considering a MF  distribution, the usual 
power-law space correlation properties correspond just
 to  a single exponent of the $f(\alpha)$ spectrum: such 
an exponent simply describes the space distribution of 
the support of the MF measure.  Furthermore the shape
 of the luminosity function (LF), i.e. the probability of 
finding a  galaxy of a certain luminosity per unit volume, 
is related to the $f(\alpha)$ spectrum of exponents of the
 MF. We have shown that, under MF conditions, the LF
 is well  approximated by a power law function with an
 exponential tail. Such a function corresponds to the 
Schechter LF observed in real galaxy catalogs. In
 this case the  shape of the LF is almost independent
 on the sample size. Indeed we have shown that a 
weak dependence on sample size   is still present 
because the cut-off of the Schechter function  
for a MF distribution turns out to be related 
to the sample  depth: $L^*$ increases with
 sample depth. In practice as this quantity
 is a strongly fluctuating one, in order to
 study its dependence on the sample size
 one should have a  very large sample 
and should vary the depth over a  
large range of length scales. Given
 this situation  a sample size independent  shape of the LF can be
 well  defined using the inhomogeneity-independent method in 
magnitude limited  samples. Indeed such a technique has been 
introduced to take into account the highly irregular  nature  of
 the large scale galaxy distribution. For example  a fractal 
distribution is non-analytic in each point and it is not possible 
to define a meaningful average density. This is because the
 intrinsic fluctuations that characterize  such a distribution 
can be large as the sample itself, and the  extent of the 
largest structures is limited only by the  boundaries of 
the available catalogs.   Moreover if the distribution is
  MF, the amplitude of the  LF depends on the sample
 size as a power law function. To determine the amplitude 
of the LF, as well as the  average density, one should have
 a well defined volume limited sample,  extracted from  
a three dimensional survey .

 In this scheme {\it the space correlations and the 
luminosity function are then two aspects of the 
same phenomenon, the MF distribution of visible matter}. 
The more complete and direct way to study such a  
distribution, and hence at the same time the space  
and the luminosity properties, is represented by the 
 computation of the MF spectrum of exponents. 
This is the natural objective  of theoretical investigation 
in order  to explain the formation and the distribution 
of galactic structures. In fact, from a theoretical point
 of view one would like  to identify the dynamical 
processes that can lead to such a  MF distribution. 

\item{\bf About Multiscaling}

Previously we have introduced the MF spectrum
 $f(\alpha)$  and now we clarify its basic properties.
 Multifractality implies that if we select only the largest peaks
 in the measure distribution, the set defined by these peaks may
 have different fractal dimension  than the set defined by the
 entire distribution. One can define a cut-off in the measure and
 consider only those singularities that are above it. If the
 distribution is MF the fractal dimension decreases as the cut-off
 increases. We note that, strictly speaking, the presence of the
 cut-off can lead (for a certain well defined value of the  cut-off
 itself) to the so-called {\it multiscaling} behavior of the MF
 measure \cite{jan91}.
In fact, the presence of a
 lower cut-off  in the calculation of the generalized correlation
 function affects the single-scaling regime of $\chi(\epsilon,q)$
 for a well determined value of the cut-off $\alpha_{cut-off}$ such
 that $\alpha_{cut-off} < \alpha_c$,  and this function exhibits a
 slowing varying exponent proportional to the  logarithm of the
 scale $\epsilon$. However some authors \cite{mar95}
misinterpret the multiscaling of a MF distribution as the
 variation of the fractal dimension with the  density of the
 sample, or with the galaxy luminosities. 

  The fractal dimension $D$ of the support corresponds to
 the  peak of the $f(\alpha)$-spectrum  and raising the cut-off
 implies a drift of $\alpha$ towards $\alpha_{min}$ so that
 $f(\alpha)<D$. This behavior can be connected with the
 different correlation exponent found by the angular correlation
 function for the elliptical, lenticular and spiral  galaxies.
In particular the observational evidence is that the
 correlation exponent is higher for elliptical than for spiral
 galaxies: this trend is compatible with a  lower fractal dimension
 for the more massive galaxies than for  the smaller ones, in
 agreement with a  MF behaviour. \bigskip

\item{\bf Problem of $\delta N/N$ and $\sigma^2$: linear and non linear dynamics}

In the discussion of large-scale
structures, is that it is true that
larger samples show larger structures
but their amplitudes are smaller
and the value of $\:\delta N / N$
tends to zero at the limits of the
sample; therefore one expects
that just going a bit further,
homogeneity would finally be
observed. Apart from the fact
that this expectation has been
systematically disproved,
the argument is conceptually
wrong for the same reasons
of the previous discussion.
In fact, we can consider a
portion of a fractal structure
of size $\:R_{s}$ and study
the behavior of  $\:\delta N / N$.
The average density $\:N$ is
just given by Eq.\ref{l5} while
the overdensity
$\:\delta N$, as a function
of the size $\:r$ of a given
in structure is ($\:r\leq R_{s}$):
\be
\label {e3n1}
\delta N = \frac{N(r)}{V(r)} - <n>
= \frac {3}{4\pi} B (r^{-(3-D)}-R_{s}^{-(3 - D)})
\ee
We have therefore
\be
\label {e3n2}
\frac {\delta N}{N} =
\left(\frac{r}{R_{s}}\right)
^{-(3 - D)} - 1
\ee
Clearly for structures that
approach the size of the small
sample, the value of  $\:\delta N / N$
becomes very small and eventually
becomes zero at $\:r = R_{s}$.

This behavior, however, could not be
interpreted as a tendency towards
homogeneity because again
the exercise refers to a self-similar
fractal by construction. Also
in this case the problems
come from the fact that
one defines an "amplitude"
arbitrary by normalizing
with the average density
that is not an intrinsic
quantity.
A clarification of this
point is very important
because the argument that
since  $\:\delta N / N$ becomes
smaller at large scale, there
is a clear evidence of
homogenization is still
quite popular \cite{pee93}
and it provides
to add confusion to the
discussion.

The correct interpretation of
 $\:\delta N / N$ is also
fundamental for the development
of the appropriate theoretical
concepts. For example a popular
point of view is to say that
 $\:\delta N / N$ is large $\:( \gg 1)$
for small structure and this
implies that a non linear
theory will be necessary to
explain this. On the other hand
 $\:\delta N / N$ becomes small
$\:(<1)$ for large structures,
which require therefore a linear
theory.
The value of  $\:\delta N / N$
has therefore generated a conceptual
distinction between small structures
that would entail non linear dynamics
and large structures with small
amplitudes that correspond instead
to a linear dynamics.
If one would apply the same
reasoning to a 
fractal structure we would conclude
that for a structure up to 
(from Eq.\ref {e3n2}):
\be
\label {e3n3}
r^{*} = 2^{-\left( \frac{1}{3-D} \right)} R_{s}
\ee
we have  $\:\delta N / N >1$ and so that a non linear
theory is needed. On the other hand, for
large structures $\:(r>r^{*})$ we have
 $\:\delta N / N < 1$ that would correspond to a
linear dynamics. Since the fractal
structure, that we have used
to make this conceptual exercise,
has scale invariant structures
by construction, we can see
the distinction between
linear and non linear dynamics
is completely artificial
and wrong.
The point is again that the value
of $\:N$, we use to normalize
the fluctuations is not
intrinsic, but it just
reflects the size of the sample
that we consider ($\:R_{s}$).

If we have a sample with  depth $\:\tilde{R_{s}}$
greater than the eventual scale of homogeneity $\:\lambda_{0}$,
then the average density will be constant in the range
$\:\lambda_{0} < r < \tilde{R_{s}}$, apart from small amplitude
fluctuations. The distance at which $\:\delta N/N =1$ will be
given by:
\be
\label {e3n4}
r^* = 2^{-\left( \frac{1}{3-D} \right)} \lambda_{0}
\ee
If, for example, $\:D=2$ and $\:\lambda_{0}=200 Mpc$ then
$\:r^{*} = 100 Mpc$:
therefore a homogeneity scale of this order of magnitude is
incompatible with the standard normalization of
$\:\delta N/N =1$ at $\:8 h^{-1} Mpc$.

We can see therefore that the whole
discussion about large and small
amplitudes and the corresponding
non linear and linear dynamics,
has no meaning until an unambiguous
value of the average density has been defined, so
that the concepts like large and small amplitudes
can take a physical meaning and be
independent on the size of the
catalogue.

The basic point of all this
discussion is that in a
self-similar structure one cannot
say that correlation are
"large" or "small", because
these words have no physical meaning due
to the lack of a characteristic
quantity with respect to which
one can normalize these properties.
The deep implication of this
fact is that one cannot
discuss a self-similar structure in terms
of amplitudes of correlation. The only
meaningful physical quantity is the
exponent that characterized the
power law behavior. Note that this
problem of the "amplitude" is not
only present in the data analysis but
also in the theoretical models.
Meaningful amplitudes can only be
defined once one has unambiguous
evidence for homogeneity but this
is clearly not 
the case for galaxy and cluster distributions.

\item{\bf Power laws, self-similarity and non analiticity:
  Amplitudes versus exponents}

\bef
\epsfxsize 12cm
\centerline{\epsfbox{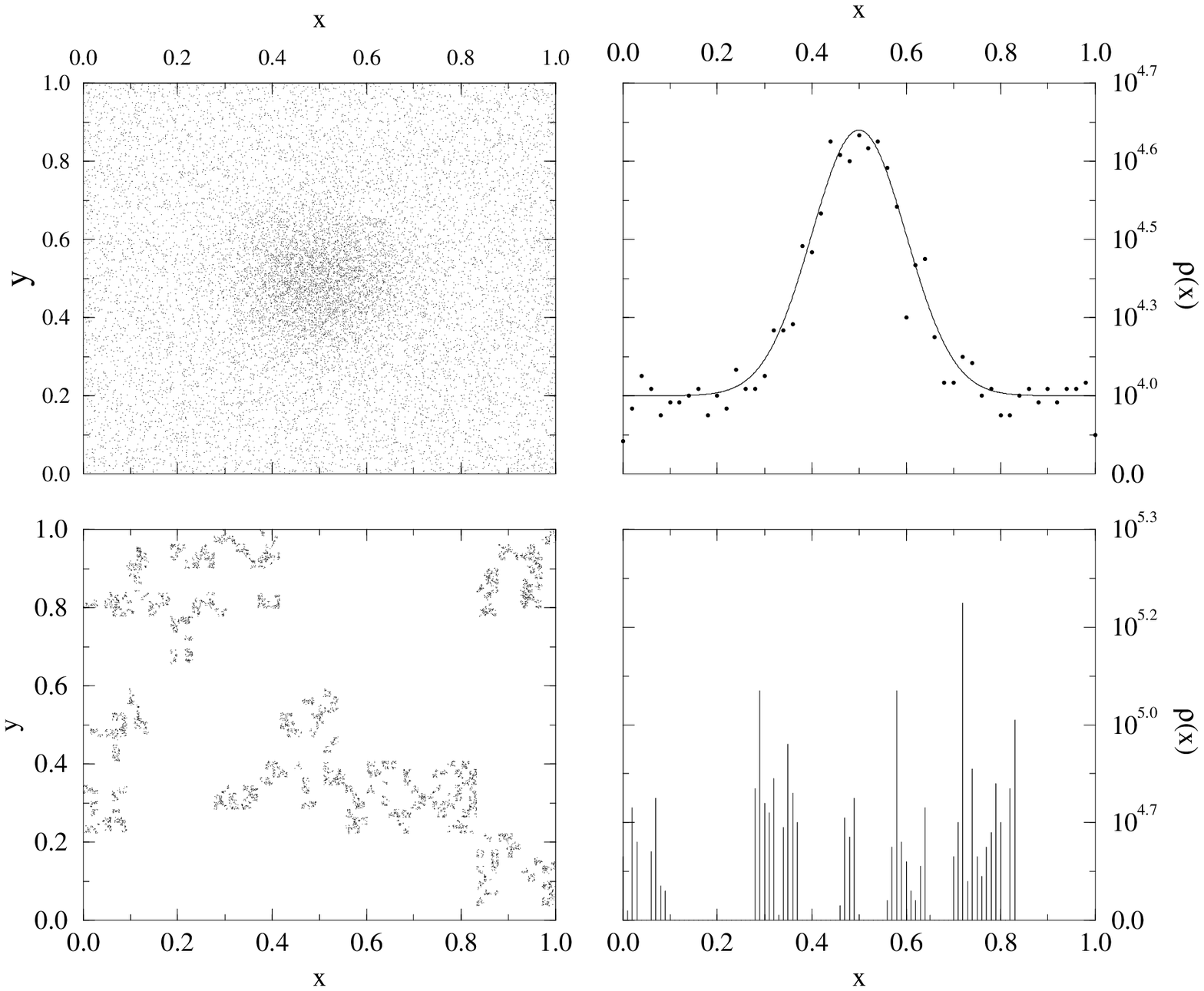}}
\caption{\label{fig4} 
Example of analytical and nonanalytic structures. {\it Top panels}
(Left)  A cluster in a homogenous distribution. (Right)
Density profile. In this case the fluctuation
 corresponds to an enhancement
of a factor 3 with respect to the average density.
{\it Bottom panels} (Left) Fractal distribution 
in the two dimensional Euclidean space. (Right) Density profile. In this 
case the fluctuations are non-analytical and there is no 
  reference value, i.e. the average density. The average density
scales as a power law from any occupied point of the structure.
}
\eef

Most of theoretical physics is based on analytical functions and 
differential equations. This implies that structures should be 
essentially smooth and irregularities are treated as single fluctuations 
or isolated singularities. The study of critical phenomena and the 
development of the Renormalization Group (RG) theory in the 
seventies was therefore a major breakthrough \cite{wil82} \cite{amit86}.
One could observe and describe phenomena in which {\it intrinsic 
self-similar irregularities develop at all scales} and fluctuations cannot 
be described in terms of analytical functions. The theoretical methods 
to describe this situation could not be based on ordinary differential 
equations because self-similarity implies the absence of analyticity 
and the familiar mathematical physics becomes inapplicable. In some 
sense the RG corresponds to the search of a space in which the 
problem becomes again analytical. This is the space of scale 
transformations but not the real space in which fluctuations are 
extremely irregular. For a while this peculiar situation seemed to be 
restricted to the specific critical point corresponding to the competition 
between order and disorder. In the past years instead, the 
development of Fractal Geometry \cite{man83},
has allowed us to 
realize that a large variety of structures in nature are intrinsically 
irregular and self-similar (Fig.\ref{fig4}). 

Mathematically this situation corresponds to the fact that these 
structures are singular in every point.  This property can be now 
characterized in a quantitative mathematical way by the fractal 
dimension and other suitable concepts. However, given these subtle 
properties, it is clear that making a theory for the physical origin of 
these structures is going to be a rather challenging task. This is 
actually the objective of the present activity in the field 
\cite{epv95}.
The main difference between the popular fractals like coastlines, 
mountains, trees, clouds, lightnings etc. and the self-similarity of 
critical phenomena is that criticality at phase transitions occurs only 
with an extremely accurate fine tuning of the critical parameters 
involved. In the more familiar structures observed in nature, instead, 
the fractal properties are self-organized, they develop spontaneously 
from the dynamical process. It is probably in view of this important 
difference that the two fields of critical phenomena and Fractal 
Geometry have proceeded somewhat independently, at least at the 
beginning.

The fact that we are traditionally accustomed to think in terms of 
analytical structures has a crucial effect of the type of questions we 
ask and on the methods we use to answer them. If one has never been 
exposed to the subtileness on nonanalytic structures, it is natural that 
analyticity is not even questioned. It is only after the above 
developments that we could realize that the property of analyticity 
can be tested experimentally and that it may or may not be present  
in a given physical system.

These results have important consequences from a 
theoretical point of view. In fact, when one deals 
with self-similar structures the relevant  physical 
phenomenon that leads to the scale-invariant 
structures is characterized by the {\it exponent} 
and {\it not the amplitude} of the  physical 
quantities that characterizes such distributions. 

Indeed, the only relevant and meaningful quantity is the  exponent 
of the power law correlation function (or of the space density), 
while the amplitude of the correlation  function, or of the space 
density and of the LF, is just  related to the sample size and to 
the lower cut-offs of the distribution.  The geometric 
self-similarity has deep implications for the  non-analyticity 
of these structures. In fact, analyticity or regularity would 
imply that at some small scale  the profile becomes smooth 
and one can define  a unique tangent. Clearly this is impossible 
in a self-similar structure because at any small scale a new 
structure appears and the  distribution is never smooth.
 Self-similar structures are therefore intrinsically irregular 
at all scales and correspondingly one  has to change the 
theoretical framework into one which is  capable of 
dealing with non-analytical fluctuations. This means
 going from differential equations to something like 
the  Renormalization Group to study the exponents.
 For example the so-called "Biased theory of galaxy
 formation" \cite{kai84}
is implemented considering 
the evolution of  density fluctuations within an analytic 
Gaussian framework,  while the non-analyticity of fractal 
fluctuations  implies a breakdown of the central limit 
theorem which is the  cornerstone of Gaussian processes 
\cite{pie87} \cite{cp92} \cite{epv95} \cite{bslmp94}.

\item{\bf Number counts and "evolution"}

  Historically \cite{hu26} \cite{pee93} 
the oldest type of data about galaxy distribution is given by
the relation between the 
number of observed galaxies $N(>f)$  and their apparent
brightness $f$. It is easy to show that \cite{pee93}
\begin{equation}
\label{counts1}
N( >f)  \sim  f^{-\frac{D}{2}}
\end{equation}
where $D$
 is the fractal dimension of the galaxy distribution. Usually this
relation is written in terms of the apparent magnitude 
$f \sim 10^{-0.4 m}$ 
(note that bright galaxies correspond to small $m$).
In terms of $m$, Eq.\ref{counts1} becomes
$\log N(<m)   \sim \alpha m$ with $\alpha = D/5$ 
\cite{slgmp96} \cite{pee93}. 
The behaviour of the number versus magnitude relation 
($N(<m)$) is reported in Fig.\ref{fig5}. 
\bef
\epsfxsize 10cm
\centerline{\epsfbox{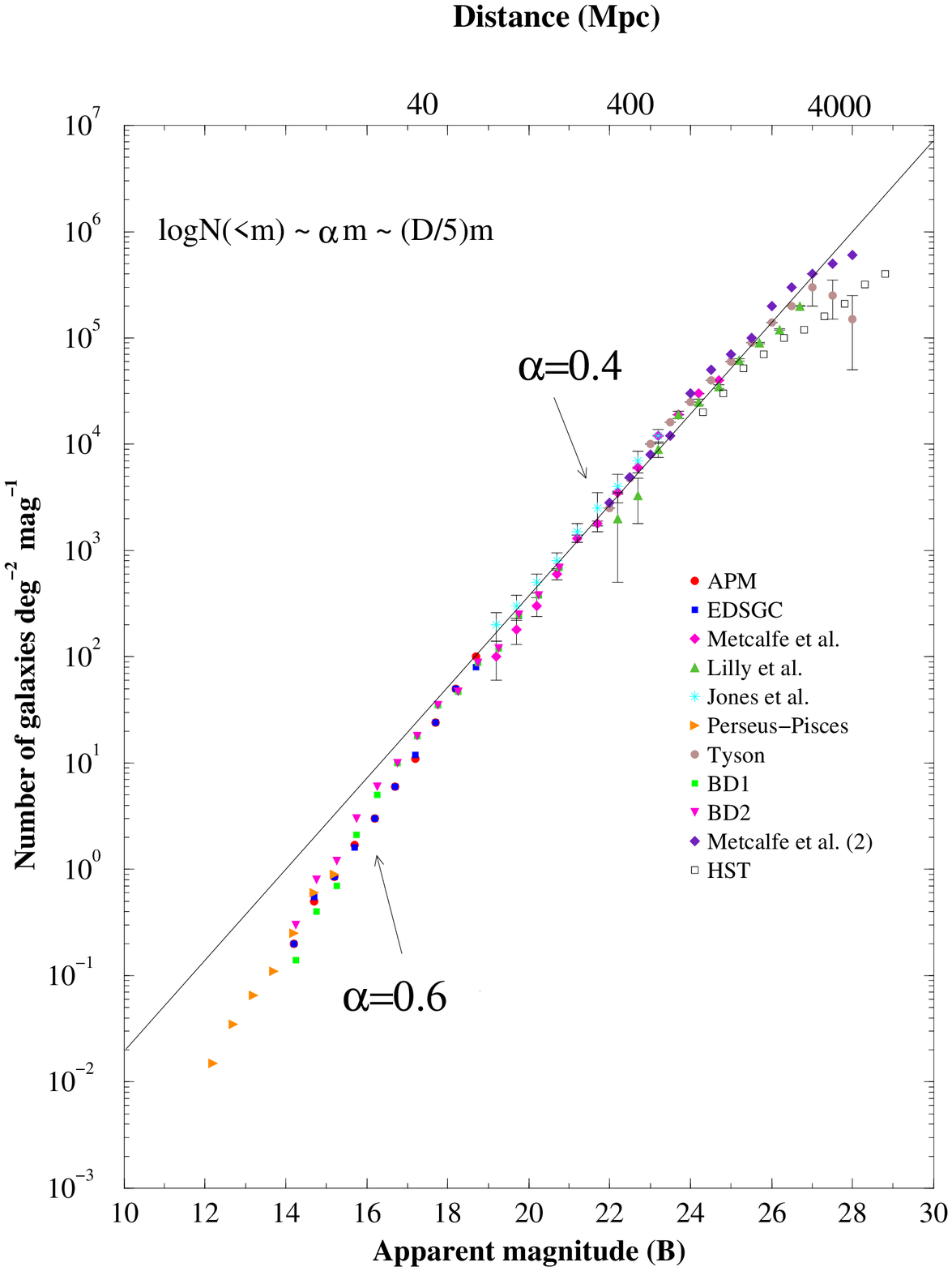}}
\caption{ \label{fig5}
The galaxy number counts in the $B$-band,
from several surveys.
In the range $12 \ltapprox  m \ltapprox 19$
the counts show an exponent 
$\alpha \simeq 0.6 \pm 0.1$, while
in the range $19 \ltapprox m \ltapprox 28$ 
the exponent is $\alpha  \simeq 0.4$.
The amplitude of the galaxy number counts for $m \gtapprox 19$
(solid line)  is computed from the determination of the 
prefactor $B$ of the density $n(r) = Br^{D-3}$ (with $D=2$ - see text) 
at small scale
and from the knowledge of the galaxy luminosity function. The distance 
is computed for a galaxy with $M=-16$ and we have used 
$H_o=75 km sec^{-1} Mpc^{-1}$.}
\eef
 One can see that at small scales (small
$m$) the exponent is $\alpha \approx 0.6$, while at larger scales 
(large $m$) it
changes into $\alpha  \approx 0.4$. The usual interpretation
\cite{pee93} \cite{yo93}  is that $\alpha \approx  0.6$
corresponds to $D \approx 3$ consistent with homogeneity, while at 
large scales
galaxy evolution and space time expansion effects are invoked to explain 
the lower value
$\alpha  \approx 0.4$.
On the basis of the previous discussion of the VL samples we can 
see that this
interpretation is untenable. In 
fact, there are very clear evidences that, at least up 
to $150  \hmp $
there are fractal correlations \cite{cp92}-\cite{dmppsl96},
 so one would eventually expect the
opposite behaviour. Namely  small value of 
$\alpha \approx 0.4$ (consistent with $D \approx 2$) at
small scales, where the effects of galaxy evolution, the K-corrections,
or the 
modification of Euclidean geometry are certainly negligible 
($z \ltapprox 0.05$) 
followed by a crossover to an eventual homogeneous 
distribution at
large scales ($\alpha  \approx 0.6$ and $D  \approx 3$).

We are going to see that this conflictual situation arises from the 
fact that,
given the limited amount of statistical information corresponding to 
the
various methods of analysis, only some of them can be considered 
as
statistically valid, while others are strongly affected by finite 
size and other 
spurious fluctuations that may be confused with real 
homogenization \cite{slgmp96}. In order to understand 
the nature of the finite size fluctuations arising in the observation from a single
point, we have to briefly discuss the case of the radial density, 
i.e. the conditional density computed from the origin.

Previously we have discussed the methods
that allow one to measure the conditional (average) density
in real galaxy surveys. This statistical quantity is an average
 one, since it is determined by making an average over all the
 points of the sample. We have discussed in detail the robustness
 and the limits of such a measurement. In particular, we have seen
 that the estimation of the conditional density can be done up to
 a distance $R_{s}$ that is of the order of the radius of the
 maximum sphere fully contained in the sample volume. This is
 because the conditional density must be computed {\it only 
in spherical shells.}
This condition puts a great limitation to the volume studied,
 especially in the case of 
deep and narrow surveys, for which the maximum depth $R_{d}$ 
can be 
one order of magnitude, or more, than the effective depth $R_{s}$.

Here we discuss the measurement of the {\it radial density} 
in VL samples \cite{slgmp96}. The determination of such a quantity
 will allow us to extend the analysis of the space density well
 beyond the depth   $R_{s}$.
The price to pay is that such a measurement is strongly affected by
 finite size spurious fluctuations, {\it because it is not an average
 quantity. } These finite size effects require a great cautious, as
 we are going to see in the following \cite{slgmp96}.

Considering  homogeneous distribution we can define, in  average,
 a characteristics volume associated to each particle. This is the
 Voronoi volume $v_v$ whose radius $\ell_v$ is of the
 order of the mean particle  separation. It is clear that the
 statistical properties of the system can be defined only in
 volumes much larger than $v_v$. Up to this volume in fact we
 observe essentially nothing. Then one begins to include a few
 (strongly fluctuating) points, and finally, the correct scaling
 behavior is recovered  (Fig.\ref{fig6}). 
\bef 
\epsfxsize 8cm 
\centerline{\epsfbox{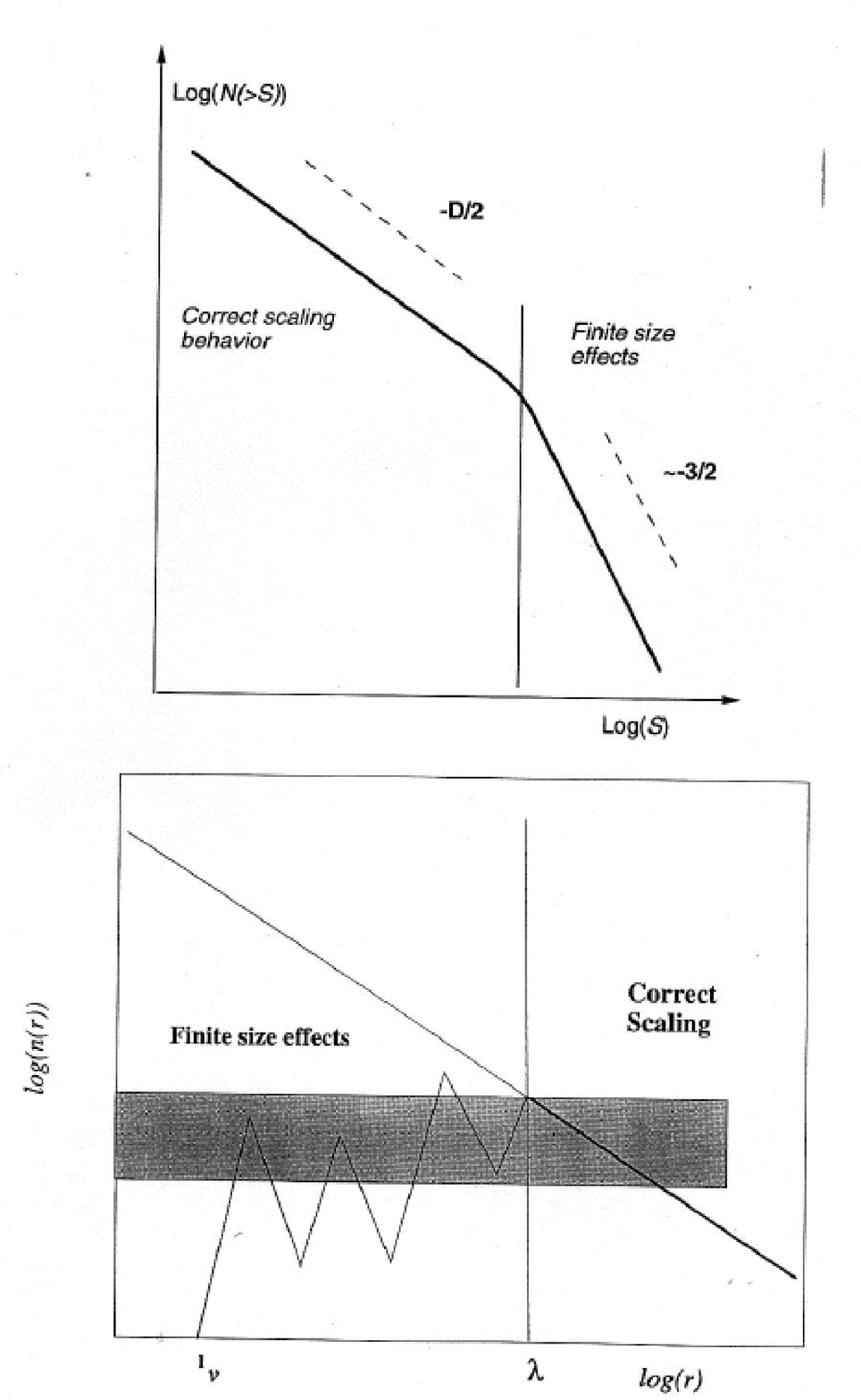}} 
\caption{\label{fig6} 
$(a)$The behaviour of the conditional density 
computed from a single point in the ideal case of a fractal
structure. Before the distance $\ell_v$
(that is of the order of the 
characteristic size of the Voronoi polyhedron) in average, 
one
 does not find any other points. Beyond this distance one sees a fluctuating 
region up to the scale $\lambda$ that is related to the 
intrinsic properties of the fractal structure. Finally
 the correct scaling regime is reached.
 $(b)$ The $N(>S)$ relation for the fractal structure
whose density is shown in $(a)$. At faint fluxes, corresponding to large
distances, one observes the correct scaling behaviour with an 
exponent $-D/2$, while at bright fluxes the finite size effects
 dominate the behaviour. 
In this case one detects an exponent $\gtapprox -3/2$ that seems to be 
in agreement with the homogenous case, but that is just due to the highly
fluctuating behaviour of the density 
 } \eef 
For a Poisson sample consisting of $N$ particles inside a volume
 $V$ then the Voronoi volume 
is of the order 
\be
 \label{v1} 
v_v \sim \frac{V}{N} 
\ee 
and $\ell_v \approx v_v^{1/3}$. In the case of homogeneous
 distribution, where the fluctuations have  {\em small amplitude}
 with respect to the average density, one  readily recovers the
 statistical properties of the system at small distances, say, $ r
 \gtapprox 5 \ell_v$. 

The case of fractal distribution is more subtle. For a self-similar
 distribution one has, within a certain radius $r_0$, $N_0$
 objects. Following \cite{cp92} we can write the mass-length
 relation between $N(<R)$, the number of points inside a sphere of
 radius $R$, and the distance $R$ of the type 
\be 
\label{new1} 
N(<R) = B R^D 
\ee 
where the prefactor $B$ is related to the lower cut-offs $N_0$ and
 $r_0$ 
\be 
\label{new2} 
B=\frac{N_0}{r_0^D} \; .
\ee 
In this case, the prefactor $B$ is defined for spherical  samples.
 If we have a portion of a sphere characterized by a solid angle
 $\Omega$, we write Eq.\ref{new1} as 
\be 
\label{new1a} 
N(<R) = B R^D \frac{\Omega}{4\pi} \;.
\ee 

In the case of a finite fractal structure, we have to take into account 
the statistical fluctuations. In the paper \cite{slgmp96} 
we have proposed an argument to take into accounts 
the finite size fluctuations based on extension of the 
concept of Voronoi length in the case of  a fractal. Such an argument 
holds for samples in which the scaling region is of the order 
of the finite size effects region.
 Here we present a new and more general argument 
for the description of the finite size fluctuations, 
that is heuristic as well (see \cite{mon97} for a more detailed 
discussion of the finite size fluctuations). 

We can identify two basic kind of fluctuations:
the first ones are intrinsic $f(r)$  and are due to
the highly fluctuating nature of fractal distributions.
Such an intrinsic noise can be seen as a modulating term 
in Eq.\ref{new1a} \cite{slmp97}.
The second $F(r)$
 term is an additive one, and it takes
 into account spurious finite size fluctuations
is simply due to shot nous 
This term becomes negligible if the shot noise fluctuations
are small: for example, if
\be
\label{cazz8}
N(<r)  > 10 \sqrt N(<r) \; .
\ee
From this condition we can have a condition on $\lambda$:
\be
\label{cazz9}
\lambda \sim \left( 10^2 \frac{4\pi}{B \Omega} \right)^{\frac{1}{D}}
\ee
The {\em minimal statistical length} $\lambda$ is an explicit
 function of the prefactor $B$ and of the solid angle
 of the survey $\Omega$.

 In the case of real galaxy catalogs we have to consider the
 luminosity selection effects. In such a case  we obtain 
for a typical volume limited sample with $M_{lim} \approx M^*$,
\be
\label{v3} 
\lambda \approx \frac{(20 \div 60) \hmp}{\Omega^{\frac{1}{D}}} \;.
\ee
where we have used $B\sim 10 \div 15 (\hmp)^{-D}$, obtained in the 
various different redhsift surveys \cite{slgmp96} \cite{slmp97}.

 We are now able to
 clarify the problem of magnitude limited (ML) catalogs. Suppose to have a certain
 survey characterized by a solid angle $\Omega$  and we ask the
 following question: up to which apparent magnitude limit 
 $m_{lim}$ we have to push  our observations to obtain that the
 majority of the galaxies lie in the statistically significant 
 region ($r \gtapprox  \lambda$) defined by Eq.\ref{v3}.  Beyond
 this value of $m_{lim}$ we should
 recover the genuine properties
 of the sample because, as we have
 enough statistics, the finite
 size effects self-average. From the previous condition for
 each $\Omega$ we can find a solid angle $m_{lim}$ 
 so that finally
 we are able to obtain $m_{lim}=m_{lim}(\Omega)$ 
in the following
 way. 

In order to give an estimation of this effect, we can 
impose the condition that, in a ML sample, the peak of 
the selection function, that occurs at distance $r_{peak}$,
satisfies the condition
\be
\label{cayy1}
r_{peak} > \lambda
\ee
where $\lambda$ in the minimal statistical length defined by Eq.\ref{v3}.
The peak of the selection function occurs for $M \approx -19$ (Sec.6) 
si that $r_{peak} \approx 10^{\frac{m_{lim}-6}{5}}$.
From the previous relation and from Eq\ref{cayy1} and Eq.\ref{v3}
we have that
\be
\label{cayy2}
\Omega \gtapprox 10^{\frac{32-2 m_{lim}}{5}}
\ee
From the previous relation it follows that for $m > 19$
 the statistically significant region  is reached for almost {\em
 any} reasonable value of the survey solid angle. This implies that
 in the deep surveys, if we have enough statistics, we   readily 
 find the right behavior ($\alpha =D/5$) while it does not happens
 in a self-averaging way for the nearby samples. Hence the exponent
 $\alpha \approx 0.4$ found in the deep surveys ($m>19$) is a {\em
 genuine feature of  galaxy distribution}, and corresponds to real
 correlation properties.  

In the nearby surveys $m < 17$ we do not
 find the scaling region in the ML sample for almost {\em any}
reasonable
 value of the solid angle. Correspondingly the value of the
 exponent is subject to the finite size effects, and to recover the
 real statistical properties of the distribution one has to perform
 an average.  

From  the previous discussion it appears now clear
 why a change of slope is found at $m \sim 19$: this is just a
 reflection of the lower cut-off of the fractal structure and in
 the surveys with $m_{lim} > 19$ the self-averaging properties of
 the distribution cancel out the finite size effects. This result
 depend very weakly on the fractal dimension $D$ and on the
 parameters of the luminosity function $\delta$ and $M^*$ used. Our
 conclusion is therefore that the exponent $\alpha \approx 0.4$ for
 $m > 19$ is a genuine feature of the galaxy distribution and it is
 related to a fractal dimension $D \approx 2$, that is found for $m
 < 19$ in redshift surveys only {\em performing averages}.  We note
 that this result is based on the assumption that the Schechter
 luminosity function  holds also at high redshift, or, at
 least   to $m \sim 20$. This result is confirmed by the analysis
 of Vettolani \etal \cite{vet94} who found that the luminosity
 function up to $z \sim 0.2$ is in excellent agreement with that
 found in local surveys \cite{dac94}.

We can now go back to Fig.\ref{fig6} and give to 
it a completely new 
interpretation. At
relatively small scales we observe 
$\alpha \approx 0.6$ just because of finite size
effects and not because of real homogeneity. This resolves the apparent
contradiction between the number counts and the 
correlation in  VL samples that show fractal
behaviour up to a few hundreds megaparsecs. In the region where 
$m>19$ we are instead sampling a
distribution in which the majority of galaxies are at distances larger 
than
$\lambda$ and indeed $\alpha  \approx 0.4$, 
corresponding to $D \approx 2$, in full agreement
with the correlation analysis. Note that the change of slope 
at $m \approx  19$ depends
only weakly on the solid angle of the survey. In order to check that 
the
exponent $\alpha  \approx 
0.4$ is the real one we have made various tests on PP where
also one observes $\alpha  \approx 
0.6$ at small values of $m$, but we know that the
sample has fractal correlations from the complete space analysis
\cite{slgmp96}. An average of
the number counts from all points leads instead to the correct exponent
$\alpha \approx  0.4$ because for average quantities the 
effective value of $\lambda$ becomes actually appreciably
smaller. We will discuss this point in detail elsewhere
\cite{slgmp96}. Our conclusion is therefore that 
there is not {\it any change of slope} at $m \sim 19$, and we 
see the same exponent in the range $ 12 \ltapprox m \ltapprox 18$,
where the combined effects K-corrections, galaxy evolution and
modification of the Euclidean geometry are certainly
negligible, and in the range 
 $ 19 \ltapprox m \ltapprox 28$. The counts, if properly determined,
do not exhibits any change of slope.

\section{Conclusions}

In summary our main points are:

\item
 The highly  irregular galaxy distributions with large structures and 
voids strongly point to a new statistical approach in which the 
existence of a well defined average density is not assumed a priori and 
the possibility of non analytical properties should be addressed 
specifically.

\item
 The new approach for the study
 of galaxy correlations in all the available catalogues 
shows that their properties are actually compatible with each other 
and they are statistically valid samples. The severe discrepancies 
between different catalogues that have led various authors to consider 
these catalogues as {\it not fair}, were due to the inappropriate methods of 
analysis.

\item
 The correct two point correlation analysis shows well defined fractal 
correlations up to the present observational limits, from 1 to 
$1000\hmp$ with fractal dimension $D \simeq 2$.
Of course the statistical quality and 
solidity of the results is stronger up to 
$100 \div 200 \hmp$ and 
weaker for larger 
scales due to the limited data. It is remarkable, 
however, that at these larger scales one observes exactly the continuation
of the correlation properties of the small and intermediate scales.

\item
 These new methods have been extended also to the analysis of the 
number counts and the angular catalogues which are shown to be fully 
compatible with the direct space correlation analysis. The new analysis of 
the number counts suggests that fractal correlations may extend also to 
scales larger that $1000\hmp.$

\item
The inclusion of the galaxy luminosity (mass) leads to a distribution 
which is shown to have well defined multifractal properties. This leads 
to a new, important relation between the luminosity function and that 
galaxy correlations in space.

\item It is worth to notice Kerscher \etal \cite{ker97} 
presented at this meeting 
 the morphological analysis of the IRAS 1.2 Jy by means of the Minkowski functional.
Their conclusion that the scale
of homogeneity is "considerably larger than $200 \hmp$", is 
in complete agreement with ours. Moreover they have done a morphological
characterization of structures that is complementary to the 
studies of the correlations properties presented in this lecture.
Finally, we would like to stress also that these authors 
find again the "apparent homogenization" due to sparse sampling: the 
same kind of effect has been discussed here and in \cite{slgmp96}.

\item 
Finally one should note that there are various
{\it indirect} arguments
and always require an interpretation based on some assumptions.
The most {\it direct} evidence for the properties of galaxy distribution 
arises from the correct correlation analysis of the 3-d volume limited samples
that has been the central point of our work.

\end{itemize}

\section*{Acknowledgments}
F.S.L. has the pleasure to thank T. Buchert and H. Wagner for
their kind hospitatlity, and the organization of the interesting discussion
we had at this meeting. 

{99}


\begin{thebibliography}{99}


\bibitem{cp92} Coleman, P.H. and Pietronero, L.,1992 Phys.Rep. 231,311


\bibitem{bslmp94} Baryshev, Y., Sylos Labini, F.,
Montuori, M., Pietronero, L. Vistas in Astron. 1994, 38, 419
     
\bibitem{pmsl97}Pietronero,L.Montuori, M.and Sylos Labini, F.
in the Proc. of the Conference "Critical Dialogues in Cosmology"
N. Turok ed., 1997

\bibitem{slmp97} Sylos Labini, F., Montuori, M.and 
 Pietronero,L.  1996, preprint



\bibitem{man83}Mandelbrot B., 1982 The Fractal Geometry of Nature,
Freeman, New York


\bibitem{pie87}Pietronero L., Physica A, 144, 257

\bibitem{cps88}Coleman, P.H.  Pietronero, L.,\& Sanders,R.H.,1988, A\&A, 245,1  


\bibitem{pee80} Peebles, P.E.J., 1980 "The Large Scale Structure of 
The Universe" 
(Princeton Univ.Press.); 


      
\bibitem{dp83} Davis, M., Peebles, P. J. E. 1983
Ap.J., 267,465



\bibitem{slgmp96} Sylos Labini, F.,
Gabrielli, A., Montuori, M., Pietronero,L.  1996, Physica A, 266, 149

\bibitem{slmp96} 
 Sylos Labini, F.,
  Montuori, M., Pietronero,L.  1996, Physica A, 230, 368;


\bibitem{slp96}Sylos Labini, F., Pietronero,L. 1996. Ap.J. 469, 28

\bibitem{sla96} Sylos Labini, F. Amendola, L. 1996, Ap.J. Lett, 468, L1

\bibitem{dmppsl96}Di Nella H.,
Montuori M., Paturel G., Pietronero L.,
and Sylos Labini F., Astron.Astrophys.Lett. 1996, 308,  L33

\bibitem{par94} Park, C., Vogeley, M.S., Geller, M., Huchra, J.
 1994
Ap.J., 431, 569;

\bibitem{ben96} Benoist C. \etal 1996 ApJ in print

\bibitem{dav88} Davis M. \etal, 1988 ApJ Lett 333, L9

\bibitem{dav97} Davis M.,  (astro-ph/9610149)
in the Proc. of the Conference "Critical Dialogues in Cosmology"
N. Turok ed., 1997


\bibitem{dac94} Da Costa L.N., \etal  , ApJ. Lett  424, (1994) L1


\bibitem{pt86} Pietronero L., and Tosatti E. Eds (1986)
Fractal in Physics 
North-Holland, Amsterdam

\bibitem{epv95} Erzan A, Pietronero L., Vespignani A.
Rev. Mod. Phys. 1995, 67, 554
       
\bibitem{sc76} Schecther P., Astrophys.J. 203, (1976) 297


\bibitem{pn91}Peacock, J.A., Nicholson, D.
Mon.Not.R.Acad.Soc  235, (1991) 307 



\bibitem{pv87} Paladin, G., Vulpiani,A.   Phys. Rep. 156, (1987) 147


\bibitem{ben84} Benzi,R., Paladin, G.,
Parisi, G.,Vulpiani, A. J.Phys. A., 17, (1984) 3251.


\bibitem{mj90} Martinez, V. T., 
Jones, B. J. T Mon. Not. R. astr. Soc., 242, .(1990) 517



\bibitem{jan91}Jensen M.H., Paladin G., Vulpiani A. Phys.Rev.Lett. 67,
 (1991) 208

\bibitem{mar95}Martinez {\it et al.}, Science 269, (1995) 1245


\bibitem{pee93} Peebles, P.E.J.
1993 "Principles of physical
Cosmology" (Princeton Univ.Press.)

\bibitem{wil82} Wilson K.G., 1974 Phys. Rep. 12, 75

\bibitem{amit86} Amit D., 1978
"Field theory, the Renormalization Group and Critical Phenomena
(Mc Graw-Hill, New York)

\bibitem{kai84}Kaiser N., Astrophys.J. . 284, (1984) L9

 

\bibitem{hu26}Hubble, E.  Astrophys.J. 
64, (1926) 321

\bibitem{yo93}Yoshii, Astrophys.J, 403, (1993) 552

\bibitem{mon97}
Montuori M., Amici A., Gabrielli A., Sylos Labini F., 
in preparation, 1997

\bibitem{vet94} Vettolani, G., {\em et al.} (1994)
 Proc. of Scloss Rindberg
workshop  Studying the Universe with Clusters
of Galaxies;

\bibitem{ker97} Kerscher M., Schmalzing J., Buchert T. 
and Wagner H., 1997, in this Proceeedings.
  
\end{thebibliography}
\end{document}